\documentclass[twoside,12pt]{article}
\usepackage{epsf,epsfig,amssymb,amsmath,graphicx,float}
\usepackage{color}

\begin{document}
\renewcommand{\thefootnote}{\fnsymbol{footnote}}

\begin{titlepage}

\begin{center}

\vspace{1cm}

{\Large {\bf Asymmetric Dark Matter and Sommerfeld Enhancement}}

\vspace{1cm}

{\bf Sujuan Qiu, Hoernisa Iminniyaz\footnote{Corresponding 
author, wrns@xju.edu.cn}, Wensheng Huo}

\vskip 0.15in
{\it
{School of Physics Science and Technology, Xinjiang University, \\
Urumqi 830017, China} \\
}

\abstract{We show that the relic density of asymmetric dark matter with
  long-range interactions is affected by the Sommerfeld enhancement. The 
annihilation cross section of asymmetric dark
  matter is enhanced due to the Sommerfeld effect. It results the relic
  density is decreased. Using the Planck date, the constraints on the 
asymmetry factor and coupling required to attain the observed dark matter 
density are obtained. We derive the 
upper bounds on the mass for asymmetric
  dark matter in $s-$wave and $p-$wave annihilation cases.  }
\end{center}
\end{titlepage}
\setcounter{footnote}{0}

\section{Introduction}
 Recently a number of models of asymmetric Dark Matter (DM) are proposed as
 an alternative of symmetric DM for which the particle and antiparticle are
 identical \cite{adm-models,Belyaev:2010kp,Petraki:2014uza}. Asymmetric DM 
models offer an effective 
explanation for the similarity of DM and baryon abundances, $\Omega_{DM} \sim
 \Omega_b$ \cite{Planck:2018vyg}. In asymmetric DM scenarios, DM is not neutral,
and self-conjugate, rather DM possesses particle-antiparticle asymmetry. In 
that case, the final abundance is determined not only by the annihilation
cross section, but also by the asymmetry factor which is the difference
between the particle and antiparticle abundances \cite{GSV,Iminniyaz:2011yp}. 
The antiparticle abundance is largely depressed to the negligible levels when 
the thermal bath temperature is much lower than the DM mass. The DM abundances 
in the presence of a particle-antiparticle asymmetry is computed in detail 
in Refs.\cite{GSV,Iminniyaz:2011yp}.

In asymmetric DM models, asymmetric DM is assumed to couple to the light 
scalar or vector mediators \cite{Petraki:2014uza,Feng:2009mn,Agrawal:2017rvu,Agrawal:2016quu,Cirelli:2016rnw}. The interaction between the asymmetric 
DM particles and antiparticles is appeared as long-range if the mediator is 
light enough. The wavefunction of DM particle and antiparticle is distorted by 
the long-range interaction. It is indeed the Sommerfeld effect 
\cite{Sommerfeld}, 
which leads the enhancement of DM annihilation rate at low velocities
\cite{Hisano:2002fk,Hisano:2003ec}. The couplings required to obtain the
observed DM density through the thermal freeze out mechanism of DM is
suppressed due to the Sommerfeld effect 
\cite{Hisano:2003ec,vonHarling:2014kha}. On the
other hand, the late-time DM annihilation signals are enhanced for the
certain set of couplings 
\cite{Pospelov:2008jd,MarchRussell:2008tu,Petraki:2016cnz,Hisano:2004ds,ArkaniHamed:2008qn,An:2016gad,An:2016kie,Bringmann:2016din}. 
Because the Sommerfeld enhancement
relies on the coupling of DM to the light force mediator, for the same mass, 
asymmetric DM needs stronger couplings than symmetric DM, there maybe rather
significant effect than symmetric DM in phenomenologically. 

In Refs.\cite{Baldes:2017gzw,Abudurusuli:2020exx}, the authors explored the 
effect of Sommerfeld 
enhancement on the relic density of asymmetric DM in detail when the light 
mediator is massless. The authors
deduced constraints on the mass and couplings required to obtain the observed 
value of DM abundance in Ref.\cite{Baldes:2017gzw}. Ref.\cite{Sulitan:2020mtg} discussed the impact of 
Sommerfeld enhancement on relic density in the case of s-wave annihilation
when the mediator mass $m_{\varphi} \neq 0$ in model independent way. They 
concluded that the abundance of asymmetric DM is decreased due to the
enhancement of the annihilation cross section. In the present work, we 
consider two minimal cases, in that scenarios, asymmetric DM 
is coupled to a light vector boson or a light scalar boson. We 
investigate the affect of Sommerfeld enhancement on
the relic density of asymmetric DM for $s-$wave and $p-$wave annihilations in
detail when the mediator mass $m_{\varphi} \neq 0$, especially the
effect of $p-$wave Sommerfeld enhancement on the freeze out of asymmetric DM. 
We extend our previous analysics in 
several ways. First, we plot the ratio of asymmetric DM antiparticle to
particle abundances with Sommerfeld enhancement as a function of the
inverse-scaled temperature. We find the antiparticle abundance is depleted
faster while the annihilation cross section is enhanced by the Sommerfeld
effect. The decrease of the ratio is significant for the larger
couplings. We calculate
the couplings required to attain the observed DM relic density as a function
of the asymmetry factor. The coupling is
larger for asymmetric case than the symmetric one with the Sommerfeld
enhancement. The final abundance is largely determined by the asymmetry. When
the asymmetry is large, the required mass bound is smaller. We also noticed that
the maximum value of mass needed to satisfy the observed value of DM abundance
is smaller than the symmetric DM. In our work, we 
ignored the effect of bound state formation on the relic density of asymmetric 
DM which affects the relic density of DM only around the
unitarity bound \cite{Cirelli:2016rnw,vonHarling:2014kha}.

The paper is arranged as following. In section 2, we discuss the impact of
Sommerfeld effect on the relic abundance of asymmetric DM for $s-$wave and
$p-$wave annihilation when the mediator mass $m_{\varphi} \neq 0$. In section
3, we find contraints on the parameter spaces as couplings, mass and asymmetry
when the annihilation cross section of asymmetric DM is modified by the 
Sommerfeld enhancement. The conclusion and summaries are in the last
section.

\section{Relic abundance of asymmetric DM including Sommerfeld enhancement}
We discuss the effect of Sommerfeld enhancement on the relic density of
asymmetric DM in this section. The number density of asymmetric DM particle 
$\chi$ (antiparticle $\bar\chi$) is evolved according to the following 
Boltzmann equation,
\begin{eqnarray} \label{eq:boltzmann_n}
\frac{{\rm d} n_{\chi(\bar\chi)}}{{\rm d}t} + 3 H n_{\chi(\bar\chi)} &=&  
- \langle \sigma v\rangle (n_{\chi} n_{\bar\chi} - 
n_{\chi,{\rm eq}} n_{\bar\chi,{\rm eq}})\,,
\end{eqnarray}
where $H = \pi T^2/M_{\rm Pl} \, \sqrt{g_*/90} $ is the expansion rate in the 
radiation dominated era, 
here $M_{\rm Pl} =2.4 \times 10^{18}$ GeV is the reduced Planck mass, and $g_*$ 
being the effective number of relativistic degrees of 
freedom. The equilibrium
number densities are $n_{\chi(\bar\chi),{\rm eq}} = g_\chi 
{\big[m T/(2 \pi) \big]}^{3/2}{\rm e}^{(-m \pm  \mu_\chi)/T}$, where
$g_{\chi}$ is the number
of intrinsic degrees of freedom of the particle. 
Here the chemical potentials $\mu_{\bar\chi} = - \mu_{\chi} $ when the
asymmetric DM is in equilibrium state. Thermally averaged 
Sommerfeld enhanced annihilation cross section is
\begin{equation} \label{eq:SE}
     \langle  \sigma v \rangle  =  a\, \langle S_s \rangle + 
                            b\, \langle v^2\,S_p \rangle\,.
\end{equation}
Here $a$ is the $s-$wave contribution when $v \to 0$ and $b$ is the $p-$wave
contribution when $s-$wave is suppressed. 
The Sommerfeld enhancement factors for 
the massive mediator in the case of $s-$wave and $p-$wave annihilations are 
given in \cite{Kamada:2020buc,Duerr:2018mbd,Feng:2010zp,Cassel:2009wt}. For 
$s-$wave annihilation, 
\begin{equation}\label{eq:Som_s}
S_s = \frac{2\pi \alpha}{v} 
   \frac{{\rm sinh}\left(\frac{6 m v}{\pi m_{\phi} }\right)}
   {{\rm cosh}\left(\frac{6 m v}{\pi m_{\phi} }\right) - 
    {\rm cos}\left(2\pi\sqrt{\frac{6 \alpha m}{\pi^2 m_{\phi}} - 
      \frac{ 9 m^2 v^2}{\pi^4 m^2_{\phi} }}\right)}\,,  
\end{equation}
where $v$ is the relative velocity of the
annihilating asymmetric DM particle $\chi$ and antiparticle $\bar\chi$, and 
$\alpha$ is the coupling strength. 
For $p-$wave annihilation, the Sommerfeld enhancement factor is 
\begin{equation}\label{eq:Som_p}
      S_p = \frac{\left(6 \alpha m/(\pi^2 m_{\phi}) -1 \right)^2 + 
           36 m^2 v^2/(\pi^4 m_{\phi}^2)} 
            { 1+ 36 m^2 v^2/(\pi^4 m_{\phi}^2)}  \, S_s\,.
\end{equation}
Then 
\begin{equation}
    \langle S_s\rangle = 
      \frac{x^{3/2}}{2 \sqrt{\pi}}\,\int^{\infty}_0 \,
          v^2 ~ e^{-\frac{x}{4} v^2}\, S_s\,{\rm d}v \, .
\end{equation}
\begin{equation}
    \langle v^2\,S_p\rangle = 
      \frac{x^{3/2}}{2 \sqrt{\pi}}\,\int^{\infty}_0 \,
          v^4 ~ e^{-\frac{x}{4} v^2}\, S_p\,{\rm d}v \, .
\end{equation}

For convenience, Eq.(\ref{eq:boltzmann_n}) can be expressed in terms
of $Y_{\chi(\bar\chi)} =n_{\chi(\bar\chi)}/s$, and $x = m/T$, here 
$s= 2 \pi^2 g_{*s}/45\, T^3$ is the entropy density, where $g_{*s}$ is the 
effective number of entropic degrees of freedom. Then 
\begin{equation} \label{eq:boltzmann_Y}
\frac{{\rm d} Y_{\chi(\bar\chi)}}{{\rm d}x} =
      - \frac{\lambda \langle \sigma v \rangle}{x^2}\,
     (Y_{\chi}~ Y_{\bar\chi} - Y_{\chi, {\rm eq}}~Y_{\bar\chi, {\rm eq}}   )\,,
\end{equation}
here the entropy conservation is used to derive Eq.(\ref{eq:boltzmann_Y}) and
$\lambda = 1.32\,m M_{\rm Pl}\, \sqrt{g_*}\,$,
 $g_*\simeq g_{*s}$ and $dg_{*s}/dx\simeq 0$.

The Boltzmann equation for particle (antiparticle) is rewritten as
\begin{equation} \label{eq:Yeta}
\frac{{\rm d} Y_{\chi(\bar\chi)}}{{\rm d}x} =
     - \frac{\lambda \langle \sigma v \rangle}{x^2}~  
     (Y_{\chi(\bar\chi)}^2 \mp \eta Y_{\chi(\bar\chi)}  - Y^2_{\rm eq})\,,
\end{equation}
here $Y_{\chi} - Y_{\bar\chi} = \eta\,$ is used which is obtained by
subtracting the Boltzmann equations for $\chi$ and $\bar\chi$ in 
Eq.(\ref{eq:boltzmann_Y}), where $\eta$ is a constant and
\begin{equation} 
Y^2_{\rm eq}= Y_{\chi,{\rm eq}} Y_{\bar\chi,{\rm
      eq}}=(0.145g_{\chi}/g_*)^2\,x^3e^{-2x}. 
\end{equation}
The Boltmann equation (\ref{eq:Yeta}) is solved following the 
standard picture of particle evolution scenarios. It is supposed 
the asymmetric DM particles and antiparticles were in thermal 
equilibrium with the standard model particles in the early universe. When the
temperature drops as $T < m$ for $m > |\mu_{\chi}|$, the interaction rate 
$\Gamma$ falls below the expansion rate $H$. At this freeze out point, 
asymmetric DM is decoupled from equilibrium state and the number densities of 
asymmetric DM in a co--moving space almost become constant
\cite{GSV,Iminniyaz:2011yp,standard-cos}. Following the method which is used
in \cite{Iminniyaz:2011yp}, the final abundance for antiparticle is determined as
\begin{equation} \label{eq:barY_cross}
Y_{\bar\chi}(x_\infty) =  \frac{\eta}{ 
  {\rm exp} \left[\, 1.32\, \eta \, m M_{\rm Pl}\,
     \sqrt{g_*} \, \int^{\infty}_{\bar{x}_F} 
     {\langle \sigma v \rangle}/ x^2 \,{\rm d}x\,\right] -1\, } \,,
\end{equation}
here
\begin{equation}
 Y_{\bar\chi}(x_\infty) \equiv \lim_{x\to \infty} Y_{\bar\chi}(x)\,.
\end{equation} 
The freeze out temperature $\bar{x}_F$ is fixed by using the standard
procedure that the freeze out occurred at the point when the deviation of the
relic abundance $Y_{\bar\chi}(\bar{x}_F)$ and 
$Y_{\bar\chi,{\rm eq}}(\bar{x}_F)$ is the same order as the equilibrium value
such as $Y_{\bar\chi}(\bar{x}_F) = (1+\xi) Y_{\bar\chi,{\rm eq}}(\bar{x}_F) $.
Here $\xi$ is a constant and usually we take 
$\xi=  \sqrt{2} -1 $ \cite{standard-cos}. 
The relic abundance for particle is obtained by using $Y_{\chi} =
Y_{\bar\chi} + \eta$, 
\begin{equation}\label{eq:Y_fin}
      Y_{\chi}(x_\infty) = \frac{\eta}
 { 1 - \exp \left[\, - 1.32\,  \eta\, m M_{\rm Pl}\,
 \sqrt{g_*} \,   \int^{\infty}_{x_F} 
\langle \sigma v \rangle/ x^2\,{\rm d}x \,   \right]}\,,
\end{equation}
where $x_F$ is the freeze out temperature for $\chi$. 
Eqs.(\ref{eq:barY_cross}) and (\ref{eq:Y_fin}) are consistent with the 
constraint $Y_{\chi} = Y_{\bar\chi} + \eta$ only if $x_F = \bar{x}_F$. The
total final DM relic density is  
\begin{eqnarray} \label{eq:omega}
 \Omega_{\rm DM}  h^2 & = & 2.76 \times 10^8\, \left[ Y_{\chi}(x_\infty) + 
                       Y_{\bar\chi}(x_\infty) \right]\,m,
\end{eqnarray}
where $\Omega_{\chi} = \rho_{\chi}/\rho_c$ with 
$\rho_{\chi}=n_{\chi} m = s_0 Y_{\chi}  $ and 
$\rho_c = 3 H^2_0 M^2_{\rm Pl}$, here $s_0 \simeq 2900$ cm$^{-3}$ is the 
present entropy density, and $H_0$ is the Hubble constant.

%

\begin{figure}[h!]
  \begin{center}
     \hspace*{-0.5cm} \includegraphics*[width=8cm]{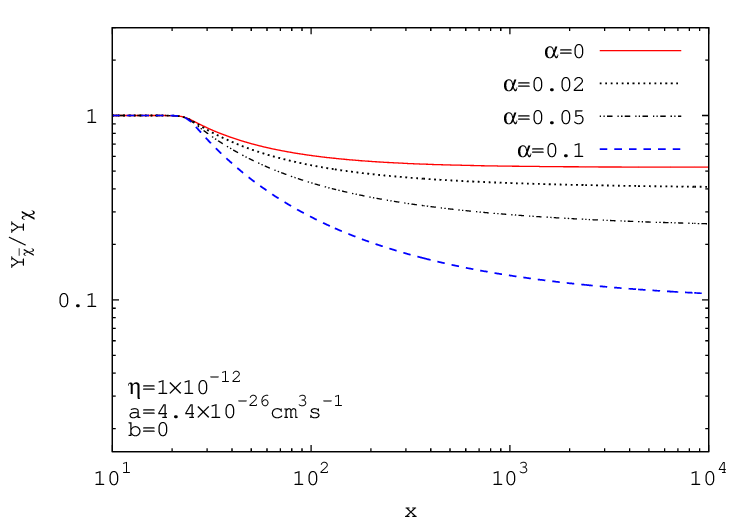}
    \put(-115,-12){(a)}
    \hspace*{-0.5cm} \includegraphics*[width=8cm]{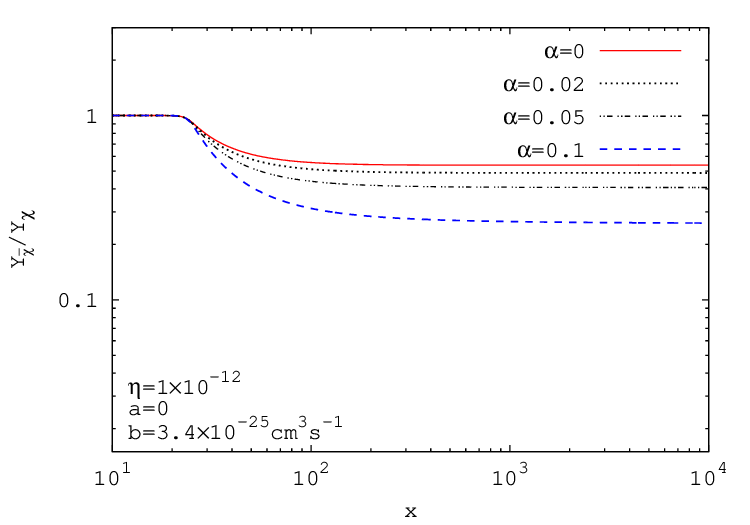}
    \put(-115,-12){(b)}
     \vspace{0.5cm}
     \hspace*{-0.5cm} \includegraphics*[width=8cm]{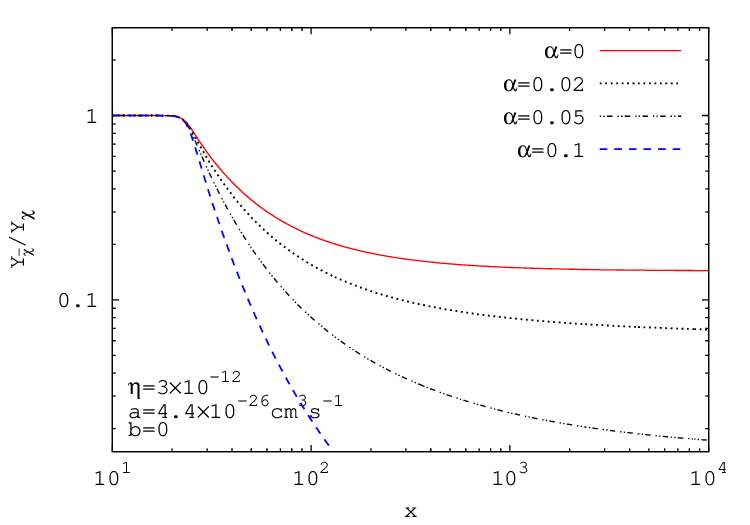}
    \put(-115,-12){(c)}
    \hspace*{-0.5cm} \includegraphics*[width=8cm]{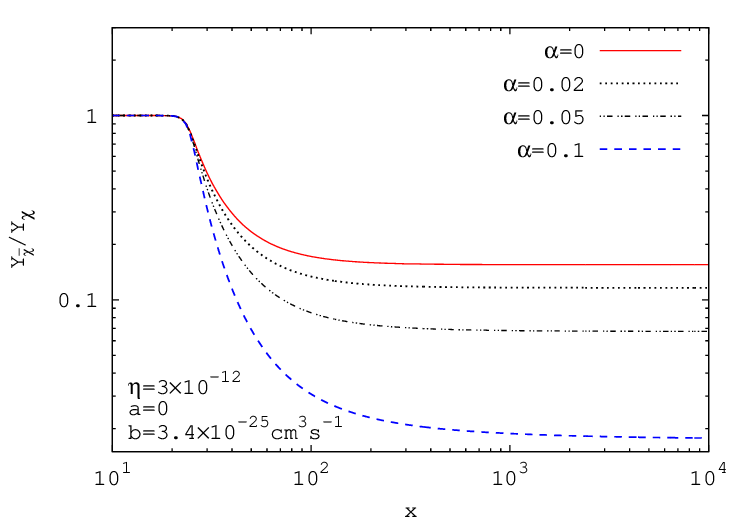}
    \put(-115,-12){(d)}
     \caption{\label{fig:a} \footnotesize
  Ratio of asymmetric DM abundances $Y_{\bar\chi}$ and $Y_{\chi}$ as a
  function of $x$ for the Sommerfeld enhanced annihilation cross section. Here 
    $m = 130$ GeV, $m_{\varphi} = 0.25$ GeV, $g_{\chi} = 2$, $g_* = 90$.}  
      \end{center}
\end{figure}

Fig. \ref{fig:a} shows the ratio of the antiparticle abundance $Y_{\bar\chi}$
to particle abundance $Y_{\chi}$ as a function of the inverse-scaled 
temperature $x$ for the Sommerfeld enhanced annihilation 
cross section. It is plotted using the numerical solutions of
Eq.(\ref{eq:Yeta}). Here panels $(a)$ and $(c)$ are for the $s-$wave
annihilations, panels $(b)$ and $(d)$ are for $p-$wave annihilations. 
The particle and antiparticle abundances keep in the same amount before the 
decoupling from equilibrium state. After freeze out, the DM abundance becomes 
almost constant as stated previously which is shown in the figure. We find the 
affect of Sommerfeld enhancement is significant for the stronger 
coupling $\alpha$ and for the larger asymmetry factor $\eta$. In other words,
the depletion of antiparticle abundance is faster when $\alpha$ is large. For 
$p-$wave annihilation, the decrease of the ratio is slower than $s-$wave
annihilation.

\section{Constraints }
In this section we discuss two minimal cases in which asymmetric DM couples to 
a light vector or scalar boson. In the case of vector mediator, we consider 
the process $\chi\bar\chi \rightarrow 2 \gamma$ which discussed in 
\cite{Baldes:2017gzw}. The Sommerfeld enhanced annihilation cross 
section times relative velocity is   
\begin{equation}\label{eq:s-wave}
    (\sigma v )_{\rm s-wave} = \sigma_0\,S_s\,,
\end{equation}
where the annihilation into two vector bosons is an $s-$wave process and the
leading order cross section is $ \sigma_0 = \pi \alpha^2/m^2$ 
\cite{Baldes:2017gzw}.

For the scalar mediator, the process $\chi \bar\chi \rightarrow 2 \varphi$ is
considered. The annihilation cross section times relative
velocity is given by
\begin{equation}\label{eq:p-wave}
    ( \sigma v )_{\rm p-wave} = \sigma_1 \,v^2 S_p\,,
\end{equation}
where the annihilation of two fermions into two scalar bosons is a $p-$wave
process and the tree level cross section is 
$\sigma_1 = 3 \pi \alpha^2/(8 m^2)$ \cite{Baldes:2017gzw}.

\begin{figure}[h]
  \begin{center}
    \hspace*{-0.5cm} \includegraphics*[width=8cm]{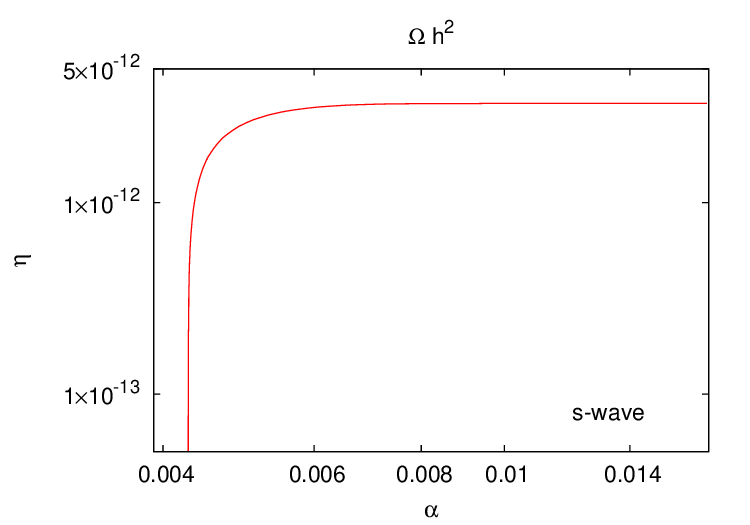}
    \put(-115,-12){(a)}
    \hspace*{-0.5cm} \includegraphics*[width=8cm]{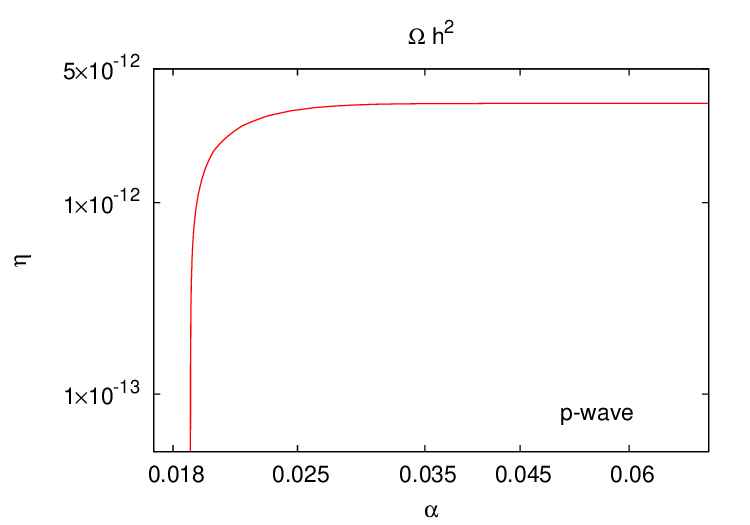}
    \put(-115,-12){(b)}
    \caption{\label{fig:b} 
    \footnotesize Contour plots of asymmetry factor $\eta$ with coupling
    strength $\alpha$ for $s-$wave ($p-$wave) annihilation cross 
    section when 
    $\Omega_{\rm DM} h^2 = 0.120$ \cite{Planck:2018vyg}. Here $m = 130$ GeV, 
    $m_{\varphi} = 0.25$ GeV, $g_{\chi} = 2$, $g_* = 90$. }
     \end{center}
\end{figure}
Following we insert Eq.(\ref{eq:s-wave}) and Eq.(\ref{eq:p-wave}) into 
Eq.(\ref{eq:Yeta}) separately and using Planck data to find constraints on the
coupling strength, asymmetry factor and mass. In Fig. \ref{fig:b}, the relation 
between asymmetry factor $\eta$ and the coupling strength $\alpha$
 for $s-$wave($p-$wave) annihilation is shown for the 
value of total DM density $\Omega_{\rm DM} = 0.120$
\cite{Planck:2018vyg}. Here $\sigma_0$
and $\sigma_1$ are evaluated using the correlated $\alpha$. The abundance is 
not sensitive for the smaller value of
$\eta$. When $\eta$ is small, the symmetric case is recovered. Therefore, the 
contour line climbs vertically (countours are independent of $\eta$ ).
On the other hand, the smaller $\eta$ gives the minimal allowed value of the 
coupling strength $\alpha =0.0043 $ for $s-$wave and $\alpha =0.0188 $ for 
$p-$wave annihilation. The lower bound of $\alpha$ is larger for $p-$wave 
annihilation. 
The curves flatten out rapidly while the asymmetry $\eta$ is 
increased. The relic abundance is determined by the asymmetry. In that case, the abundance is not affected when the coupling
strength is increased (independent of $\alpha$). A larger $\alpha$, 
corresponds to the larger
enhancement of the cross section. It leads to a smaller relic density. This
should be compensated by increasing $\eta$.

\begin{figure}[h]
  \begin{center}
    \hspace*{-0.5cm} \includegraphics*[width=8cm]{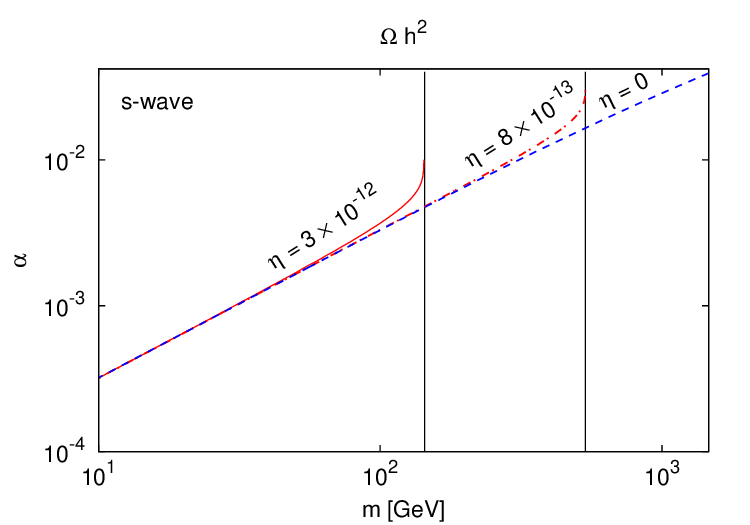}
    \put(-115,-12){(a)}
    \hspace*{-0.5cm} \includegraphics*[width=8cm]{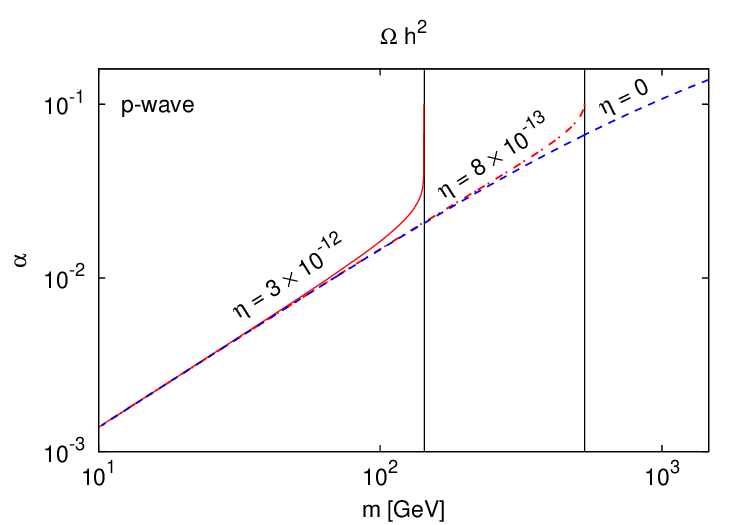}
    \put(-115,-12){(b)}
    \caption{\label{fig:c} 
    \footnotesize Contour plots of the coupling strength $\alpha$ and 
    mass $m$ for $s-$wave ($p-$wave) annihilation when 
    $\Omega_{\rm DM} h^2 = 0.120$ \cite{Planck:2018vyg}. Here  
$m_{\varphi} = 0.25$ GeV,
    $g_{\chi} = 2$, $g_* = 90$, where $\sigma_0$ and $\sigma_1$ are evaluated 
   using the correlated $\alpha$.}
     \end{center}
\end{figure}

The coupling strength required to obtain the observed value of DM relic
density with mass is plotted in Fig. \ref{fig:c}. The mass bounds for 
$\eta = 3 \times 10^{-12}\,,\,\, 8\times 10^{-13} $ are 
$m = 143\,,\,\, 535$ GeV respectively in panel $(a)$; 
$m = 143\,,\,\, 532$ GeV in panel $(b)$. Highly asymmetric DM 
has lower mass bound.  
For the same mass bound, asymmetric DM needs larger 
coupling to attain the observed value of DM relic density compared to the 
symmetric case. For example, $m = 143$, the coupling $\alpha = 0.01$ for 
$\eta = 3 \times 10^{-12}$ and $\alpha = 0.005$ for $\eta = 0$.   
The coupling for $s-$wave annihilation is smaller than the case of $p-$wave 
annihilation.  

\begin{figure}[h]
  \begin{center}
    \hspace*{-0.5cm} \includegraphics*[width=8cm]{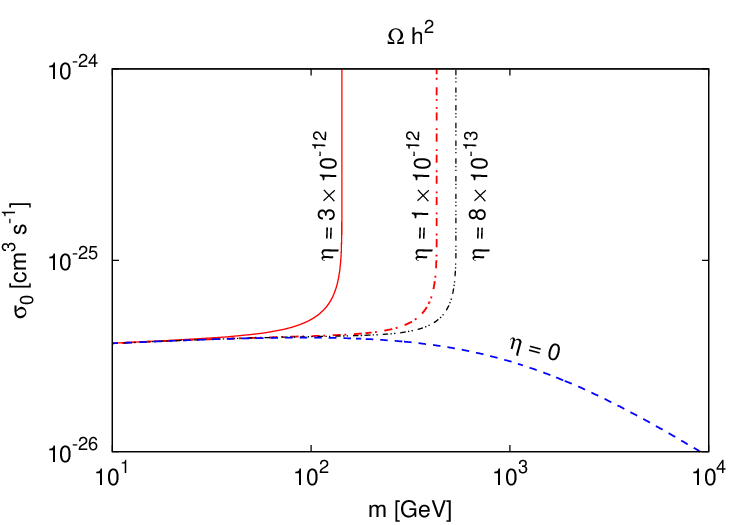}
    \put(-115,-12){(a)}
    \hspace*{-0.5cm} \includegraphics*[width=8cm]{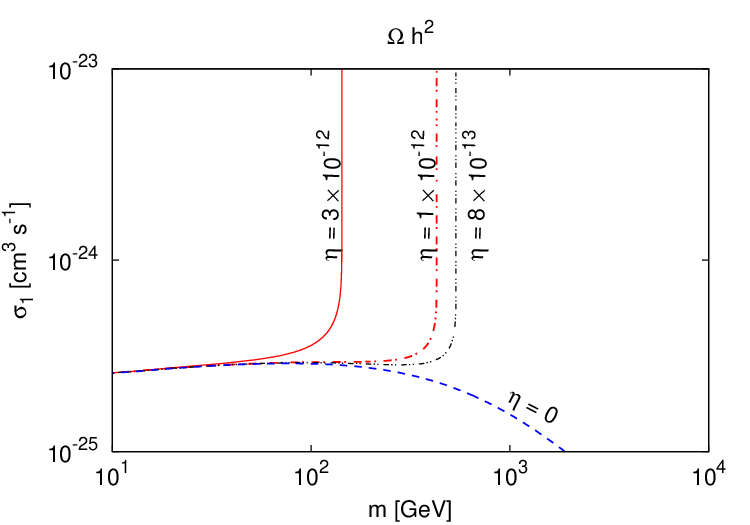}
    \put(-115,-12){(b)}
    \caption{\label{fig:d} 
    \footnotesize Contour plots of $s-$wave ($p-$wave) annihilation cross 
section $\sigma_0(\sigma_1)$ and the mass when 
    $\Omega_{\rm DM} h^2 = 0.120$. Here
    $m_{\phi} = 0.25$ GeV,
    $g_{\chi} = 2$, $g_* = 90$, the coupling $\alpha$ is computed using the 
expression
 $ \sigma_0 = \pi \alpha^2/m^2$ for $s-$wave and 
$\sigma_1 = 3 \pi \alpha^2/(8 m^2)$ for $p-$wave annihilations. }
     \end{center}
\end{figure}

The contour plots of the cross section with mass are shown in 
Fig. \ref{fig:d}. The allowed region is bounded by the cross section from below 
and by the maximum value of $m_{\rm max}$ from the right for different asymmetry.
For $m \ll m_{\rm max}$, coupling constant $\alpha$ and $\sigma_0(\sigma_1)$ 
trace the symmetric DM ($\eta = 0$) case
closely. When the cross section is small, the abundance is insensitive
for the increased mass. On the other hand, 
while $\alpha$ is small, the effect of Sommerfeld enhancement is not
significant for both the symmetric and asymmetric DM. When the mass 
reached the maximum value $m \simeq m_{\rm max}$, the Sommerfeld
enhanced annihilation cross section is increased to satisfy the observed value
of the DM relic density for asymmetric DM. In the symmetric case, for larger 
$m$, the coupling 
is stronger (see Fig. \ref{fig:c}), then the Sommerfeld enhanced cross section 
reduces the relic density notably, the cross section
should be small in order to obtain the required value of the relic density. 
Therefore, the cross section falls quickly when the mass is increased in
symmetric case.

\section{Summary and conclusions}
When the asymmetric DM coupled to the light force carrier, the annihilation
cross section is enhanced by the Sommerfeld effect. 
We discuss the effect of Sommerfeld enhancement on the relic density of
asymmetric DM for $s-$wave and $p-$wave annihilations for the case of
light mediator $m_{\varphi} \neq 0$ in detail. We found the antiparticle
abundance is depleted faster than the standard case due to the Sommerfeld 
enhanced annihilation cross section. The decrease of the ratio of antiparticle
abundance to particle abundance is larger for the stronger coupling $\alpha$. 

We apply our method to two kinds of scenarios, asymmetric DM coupled to either
light vector mediator or the scalar mediator. We use Planck data to find
constraints on the coupling constant $\alpha$, asymmetry factor $\eta$ and the 
mass value. We found when $\eta$ is small, the abundance is independent of 
$\eta$, it is indeed the symmetric case. The minimal value of the 
coupling strength are $\alpha =0.0043 $ for $s-$wave and $\alpha =0.0188 $ for 
$p-$wave annihilation. When $\eta$ takes larger value, the relic abundance is 
determined by the asymmetry and it is independent of $\alpha$.
For the same mass bound $m_{\rm max}$, asymmetric DM needs larger 
coupling to attain the observed value of DM relic density compared to the 
symmetric case. When the cross section is small, the abundance is insensitive
for the increased mass. When the mass 
reached the maximum value, the Sommerfeld
enhanced annihilation cross section is increased to satisfy the observed value
of the DM relic density for different asymmetry.

Our results are important when the affect of Sommerfeld enhancement is
significant at low
velocity limit. The Sommerfeld effect hints notable indirect detection
signals from asymmetric DM antiparticle. This allows us to examine the
asymmetric DM by the Cosmic Microwave Background (CMB) observation, the
Milky ways and Dwarf galaxies.

\section*{Acknowledgments}

The work is supported by the National Natural Science Foundation of China
(U2031204, 11765021) and Natural Science Foundation of Xinjiang Uygur
Autonomous Region (2022D01C52).

\end{document}